%% file: P12_1110_BressanA.tex
\documentclass{article}
\usepackage{amsmath,amssymb} 
\usepackage{graphicx}
\usepackage{psfrag} 
\usepackage{times} 
\begin{document}

\input{standard.tex}
\newcommand{\etal}{ et al.}
\def\Journal#1#2#3#4{{#1} {#2} (#4) #3 }
\def\NCA{{\em Nuovo Cimento} A}
\def\PHYS{{\em Physica}}
\def\NPA{{\em Nucl. Phys.} A}
\def\MATH{{\em J. Math. Phys.}}
\def\PRO{{\em Prog. Theor. Phys.}}
\def\NPB{{\em Nucl. Phys.} B}
\def\PLA{{\em Phys. Lett.} A}
\def\PLB{{\em Phys. Lett.} B}
\def\PLD{{\em Phys. Lett.} D}
\def\PL{{\em Phys. Lett.}}
\def\PRL{\em Phys. Rev. Lett.}
\def\PREV{\em Phys. Rev.}
\def\PREP{\em Phys. Rep.}
\def\PRA{{\em Phys. Rev.} A}
\def\PRD{{\em Phys. Rev.} D}
\def\PRC{{\em Phys. Rev.} C}
\def\PRB{{\em Phys. Rev.} B}
\def\ZPC{{\em Z. Phys.} C}
\def\ZPA{{\em Z. Phys.} A}
\def\ANNP{\em Ann. Phys. (N.Y.)}
\def\RMP{{\em Rev. Mod. Phys.}}
\def\CHEM{{\em J. Chem. Phys.}}
\def\INT{{\em Int. J. Mod. Phys.} E}
\def\NIM{{\em Nucl. Instr. Meth..} A}
\def\IEE{{\em IEEE Trans.Nucl.Sci.}}
%%%%%%%%%
\newcommand{\DGG}{\mrf{\Delta G/G}}
\newcommand{\DG}{\mrf{\Delta G}}
\newcommand{\dtq}{\mrf{\Delta_T q(x)}}
\newcommand{\gom}{\mrf{GeV/\mit{c}}}
%%%%%%%%%
\title{Physics Results from COMPASS}
\author{A.~Bressan\footnote{Plenary talk at the ``16\mrf{^{th}} International
    Spin Physics Symposium'', October 10-16, 2004, Trieste, Italy; to be
    published in the Conference Proceedings, World Scientific.} \\University
  of Trieste and INFN, Sezione di Trieste,\\ Trieste, Italy 
\\[2ex] on behalf of the COMPASS Collaboration}
\maketitle

\begin{abstract}The COMPASS Experiment at the CERN SPS has a broad physics program
focused on the nucleon spin structure and on hadron spectroscopy, using muon and
hadron beams.  Main objectives for the spin program with the muon beam are the
direct measurement of the gluon contribution to the spin of the nucleon,
semi-inclusive measurements, and the measurement of the transverse spin
distribution \dtq. The COMPASS apparatus consists of a two-stage large
acceptance spectrometer designed for high data rates and equipped with
high-resolution tracking, particle identification and electromagnetic and
hadronic calorimetry.\\ The data taking is ongoing since 2002 and till now was
mainly devoted to the spin programme using a 160~\gom\ naturally polarized,
$\mu^+$ beam and a polarized \mrf{^6LiD} target. First physics results from
the 2002 and 2003 runs are presented.\end{abstract}
%------------------------{\bf Introduction}---------------------------
\section{Introduction}
\label{sec:introduction}
The COMPASS\cite{k:prop} experiment is focused to a deeper understanding on how
the constituents contribute to the spin of the nucleon. The main goal is a
direct measurement of the gluon polarization \DGG,
obtained by measuring the spin dependent asymmetry of open charm production in
the photon-gluon process. The determination of the transversity distribution
function \dtq\ and studies of transverse spin effects; accurate measurements of
the flavor decomposition of the quark helicity distributions, vector meson
exclusive production and $\Lambda$ physics are also important parts of the program.
The hadron spectroscopy is dedicated to the measurements of the mass and decay
patterns of light hadronic systems and leptonic decays of charmed mesons, as
well as $\pi$ and K polarizabilities (Primakoff reactions), extensive
meson spectroscopy to investigate the presence of exotics states.

The experiment, performed by a collaboration of about 270 physicists from 27
institutes and 11 countries, was set-up in 1998--2000 and a technical run
took place in 2001. The runs from 2002 to 2004 were mainly devoted to the spin
programme with a polarized muon beam and a polarized \mrf{^6LiD} target. After
the 2005 technical stop of CERN, COMPASS will run at the SPS at least until
2010.

The talk is focused on the results of 2002 and  2003
runs, during which a total of 500 TB of data have been collected. 
%------------------------------{\bf Apparatus}------------------------------
\section{\boldmath The COMPASS apparatus}
\label{sec:setup}
The COMPASS experiment has been set up at the CERN SPS M2 beam line.
It combines high rate beams with a modern two stage
magnetic spectrometer\cite{gerd}.  Both stages are equipped with
hadronic calorimetry and muon identification via
filtering through thick absorbers. In the first stage a RICH detector is also
installed, allowing the identification of charge hadrons up to 40 GeV.  Detectors,
electronics and data acquisition system are able to handle beam rates up to
10$^8$ muons/s and about 5$\times$10$^7$ hadrons/s with a maximal interaction
rate of about 2$\times$10$^6$/s.  The triggering system and the tracking system
of COMPASS have been designed to stand the associated rate of secondaries, and
use state-of-the-art detectors.  Also, fast front-end electronics,
multi-buffering, and a large and fast storage of events are essential.

\begin{figure} 
\includegraphics[width=\textwidth]{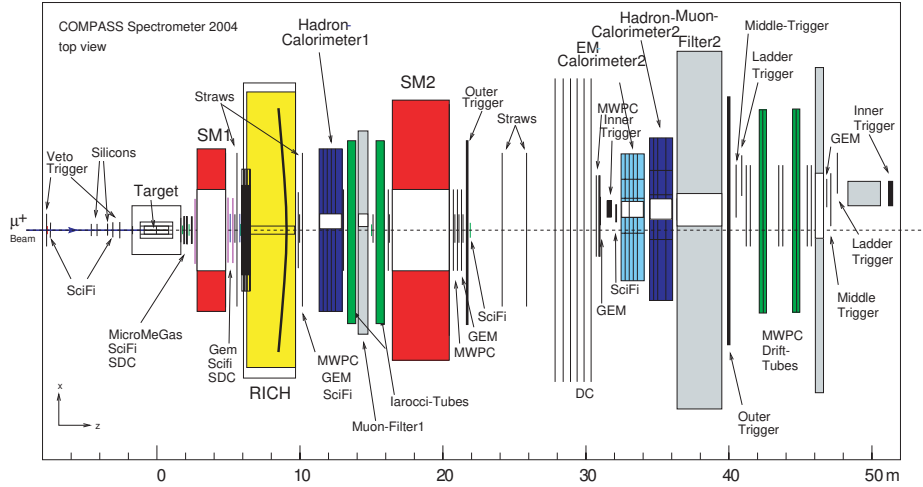} \vspace{-2ex}
\caption{Top view of the lay-out of the spectrometer for the COMPASS experiment
in 2004.  The labels and the arrows refer to the major components of the
tracking, trigger, and PID systems.}
\label{fig:setup}\end{figure}

The layout of the spectrometer for the 2004 running period is shown in
Fig.~\ref{fig:setup}.
The experiment has been run at a muon energy of 160 GeV.  The beam is naturally
polarized by the $\pi -$decay mechanism, and the beam polarization is estimated
to be 76\%.  The beam intensity is $2\times 10^8$ muons per spill (4.5 s long).

We use the polarized target system of the SMC experiment\cite{fabrice}, which allows for two
oppositely polarized target cells, 60~cm long each.  The PT magnet can provide
both a solenoid field (2.5 T), to hold the longitudinal (with respect to the
beam direction) polarization, and a
dipole field (0.5 T), needed for adiabatic spin rotation and for holding the
transverse polarization.  
Use of two different target materials, NH$_3$ as proton target
and $^6$LiD as deuteron target, is foreseen.  Polarizations of 85\% and 50\%
have been reached, respectively.  In so far we have used $^6$LiD because  its
favorable dilution factor\cite{lid} of $>$0.4 is particularly important for the
measurement of \DGG. 
In 2006 the installation of a new PT magnet build by Oxford-Danfysik, with an
increased inner bore radius matching the full acceptance of the spectrometer
($\pm 180$ mrad), is foreseen.
As a result, an increased acceptance at large \xbj\ and a better sensitivity to
the charm channel will improve the figure of merit of the experiment.\\
Different tracking devices are used to cope with different fluxes and to fit the
needed resolution. The small area trackers (SAS), sitting on or close to the
bean line, consist of several stations of scintillating fibers, silicon
detectors, micro-pattern gaseous detectors like Micromega\cite{microo}
and GEMs\cite{gems}.  Large area tracking (LAS) devices are made from
gaseous detectors (the Saclay Drift Chambers, Straw tubes\cite{straw}, MWPCs,
and the W4/5 Drift Chambers) placed around the two spectrometer magnets.
Table~\ref{tab:res02} summarizes the spatial resolution and the timing
properties of the tracking detectors, as derived from the 2002 data.
\begin{table}
\caption{Trackers performances in the 2002 run. \label{tab:res02}}
\begin{center}
{\begin{tabular}{|l|c|c|c|c|}
\hline
Detector & coordinates & efficiency & resolution & timing \\
\hline
Scintillating fibers  & 21 & 94 \%       & 130 $\mu$m  & 0.45 ns \\
Micromegas            & 12 & 95 - 98 \%  &  65 $\mu$m  & 8 ns \\
GEM                   & 40 & 95 - 98 \%  &  50 $\mu$m  & 12 ns \\
SDC                   & 24 & 94 - 97 \%  & 170 $\mu$m  &  \\
Straw tubes           & 18 & $>$ 90 \%     & ~270 $\mu$m &  \\
MWPC                  & 32 & 97 - 99 \%  &             &  \\
W4/5                  &  8 & $>$ 80 \%     &             &  \\
\hline
\end{tabular}}
\end{center}
%\caption{Trackers performances in the 2002 run.}
%\label{tab:res02}
\end{table}
Muons are efficiently identified by large detector planes placed before and
after a 60 cm thick iron absorber. Aluminum Iarocci-type limited streamer
and drift tubes planes are used in the LAS and in the SAS respectively.

Hadron identification in the LAS is provided  by RICH-1\cite{rich1}, designed
to separate $\pi$ and  K, over the whole LAS angular acceptance up to 60 GeV.
RICH-1 consists of a 3 m long \cfft\ radiator at atmospheric pressure, a wall of
spherical mirrors (3.3 focal length) covering an area of $>$20~m$^2$ and two
sets of far UV photon detectors placed above and below the acceptance region.
The Cherenkov photons are detected by MWPCs equipped with CsI photo-cathodes
\cite{RD26}, segmented in 83000, $8\times 8$ mm$^2$ pads read-out by a system of
front-end boards \cite{k:bora} with local intelligence.

The trigger is formed by two hadron calorimeters and several hodoscope systems.
2 ns wide coincidences between more than 500 elements select the scattered
muons in the kinematics region of interest, on the bases of target pointing and
energy release. An hadron shower in the hadronic calorimeter provides more
selective triggering. The overall typical trigger rate was 5 kHz with a dead
time of about 7\%. The acceptance in $Q^2$ goes from quasi-real
photons $\sim 10^{-4}$ \mrf{(GeV/c)^2} up to $\sim 100$ \mrf{(GeV/c)^2}, while
\xbj\ is from $10^{-4} (\simeq 4\times 10^{-3}\,\mrf{for\, Q^2>1}) < \xbj < 0.7$.

The readout system\cite{daq} uses a modern concept, involving highly specialized
integrated circuits.  The readout chips are placed close to the detectors and
the data are concentrated at a very early stage via high speed serial links. At
the next level high bandwidth optical links transport the data to a system of
readout buffers.  The event building is based on PCs and Gigabit or Fast
Ethernet switches and is highly scalable.  This high performance network is also
used to transfer data, via optical link, to the Central Data Recording (CDR) in
the computer center for database formatting, reconstruction, analysis and mass
storage.

The computing power needed to process the huge amount of data ($\sim$300
TB/year) is about 100 kSI2k.   
The raw data processing is centrally performed at CERN, while the DST and mDST
analysis, as well as the large Monte Carlo production are done on satellite
farms in the major home institutes.
The event reconstruction is performed by a fully object
oriented program, with a modular architecture, written in {\sf C++} (CORAL). 
{\sf C++} has also been used to write the analysis program PHAST, while
the Monte Carlo program COMGEANT is based on GEANT3.
%------------------------------{\bf 2002 Run}------------------------------
\section{First Physics Results}
\label{sec:analysis}
Many physics channels are under investigation, and important flavors of the
ongoing work were given in the parallel sessions:\\
-- $A_1^d$, the virtual photon asymmetry in both inclusive and semi-inclusive DIS\cite{dimitri},\\
-- \DGG\ from open charm and from production of pair of high-$p_T$ hadrons\cite{christian},\\
-- transverse spin asymmetries\cite{paolo,rainer},\\
-- exclusive vector meson production to test s-channel helicity conservation\cite{damien},\\
-- spontaneous $\Lambda$ polarizations\cite{vadim,jan}.\\
Due to the limited space, in this contribution only some of the relevant results
of items 1--3 will be summarized. 
\subsection{Inclusive and Semi-Inclusive Asymmetries}
\begin{figure}\begin{center}
\includegraphics[width=\columnwidth]{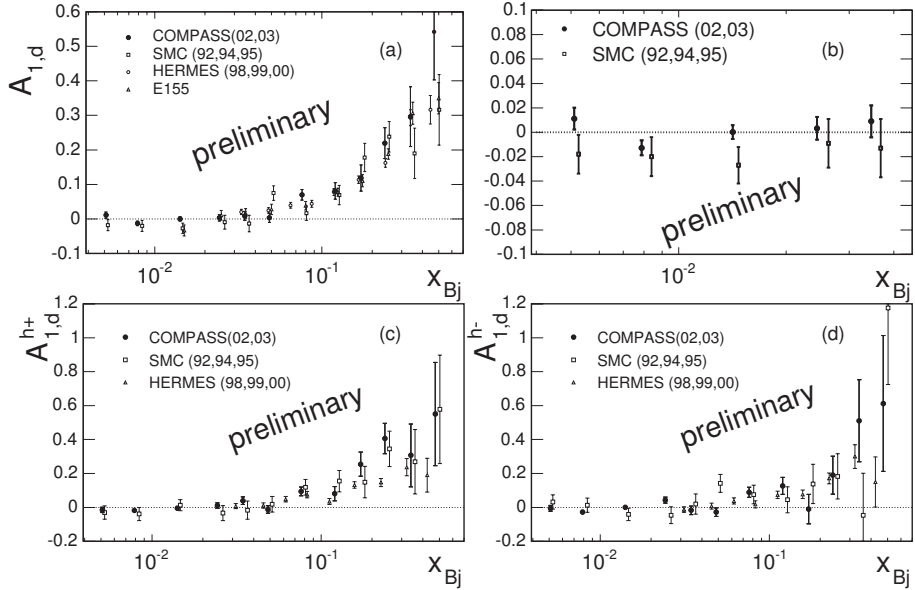} 
\end{center}\vspace{-2ex}
\caption{(a) Measured asymmetry $A_1^d$ from COMPASS 2002-2003 runs compared
  with the existing data. (b) Same measurement for the low-$x$ region compared
  with SMC data only. (c) Semi-inclusive spin asymmetry $A^h_1$ as a function of
  $x$ for positively charged hadrons. (d) Semi-inclusive spin asymmetry $A^h_1$ as a function of
  $x$ for negatively charged hadrons.}
\label{f:A1}\end{figure}

The preliminary results from COMPASS, based on 2002-2003 data, are shown in
figure~\ref{f:A1}, where results from previous measurements are also
given\cite{smc_in,slac_in,hermes_in}.  The COMPASS data are in good agreement
with the other data sets, and are compatible with zero in the low-$x$ region
(fig.~\ref{f:A1}(a) and (b)). Moreover, the COMPASS data at low-$x$
(fig.~\ref{f:A1}(b)) have statistical errors already smaller than SMC, thanks to
higher luminosity and dilution factor, and will contribute to improve the
precision on the spin dependent structure function $g_1^d(x)$ and on its first
moment $\Gamma^d_1$. 
Figures~\ref{f:A1}(c) and (d) show semi-inclusive asymmetries $A_1^h$ for
hadrons of positive charge (c) and of negative charge (d), compared with
existing data\cite{smc_si,hermes_si}. In this case statistical errors at low-$x$
are also smaller than previous measurements. Those hadrons samples are dominated
by pions ($\sim 80\%$), and at this stage of the analysis the RICH information
is not yet used. Even by using PID a full flavor decomposition from the deuteron data alone is
not possible and there is the needs to extent the measurement by collecting data
also with a proton target; nevertheless one can extract $\Delta u + \Delta d$,
$\Delta \bar{u} + \Delta \bar{d}$ and from kaon samples $\Delta s = \Delta \bar{s}$,
i.e. the the strange quark helicity distribution function. With the accumulated
statistics from 2002 to 2004, COMPASS will extend by one order of magnitude, in
the low-$x$ region, the existing results\cite{hermes-ds} on $\Delta s(x)$,
decreasing the uncertainties on the first moment.
\subsection{Gluon polarization $\Delta G/G$}
In COMPASS the gluon polarization \DGG\ is accessed by identifying the
photon-gluon fusion process, tagged either by open-charm production or by
the production of pair of high-$p_T$ hadrons.

\begin{figure}\begin{center}
\includegraphics[width=\columnwidth]{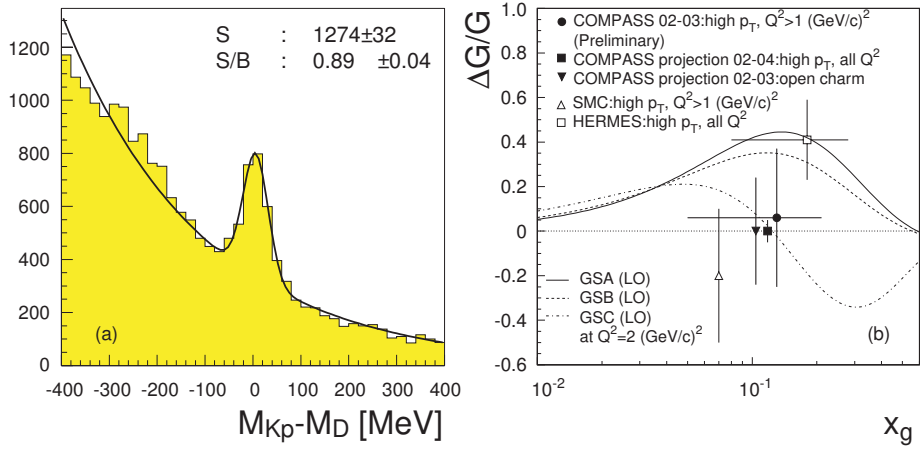}
\end{center}\vspace{-2ex}
\caption{(a) Reconstructed \mrf{D^0} in the \mrf{K \pi} invariant mass
  distribution tagged from the \mrf{D^*} decay. (b) Preliminary result for
  \DG\ from COMPASS high-$p_T$ production and 
  $Q^2 >1$ (GeV$/c$)$^2$, together with the results from HERMES ($\langle
  Q^2\rangle = 0.06$ (GeV$/c$)$^2$) and SMC ($Q^2 >1$ (GeV$/c$)$^2$). Also
  shown are the projected error bars for the open charm high-$p_T$ (all $Q^2$)
  for the whole 2002--2004 data. The curves show the parametrization A--C of
  Ref.\protect\cite{gehr-ster}.} 
\label{f:dg}\end{figure}

Open-charm events are selected by
reconstructing \mrf{D^0} and \mrf{D^*} mesons from their decay products, i.e. \mrf{D^0
\rightarrow K \pi} and \mrf{D^* \rightarrow D^0 \pi^0 \rightarrow K \pi \pi^0}.  In
the first case, cuts on the K direction in the \mrf{D^0} rest frame ($| \cos (\mrf{
\theta^*_K}) |$ $< 0.5$) and on the \mrf{D^0} energy fraction (\mrf{z_D=}
\mrf{E_D/E_{\gamma^*}} \mrf{>0.25}) are needed to reduce the background still
dominant. The second case is much cleaner given the unique kinematics. 
Figure~\ref{f:dg}(a) shows the \mrf{D^0} peak reconstructed
from the 2003 run, by selecting \mrf{D^0} coming from \mrf{D^*} decays. 
The projected error from the open charm, including all the 2002--2004 data is
$\delta \left( \DGG \right) = 0.24$. 

\DGG\ from pair of high-$p_T$ hadrons has, compared to open charm, the advantage of
a larger statistics, even if the extraction of the gluon polarization is more
difficult due to the contribution to the asymmetry of competitive processes.
Selecting events with $Q^2 > 1$ (GeV$/c$)$^2$ drastically cut the contribution from resolved
photons, at the price of a $\sim 1/10$ reduction of the data sample. 
Selecting events with high-$p_T$ hadrons reduces the contribution from the
leading order  process $\gamma q \rightarrow q$ (LO), and increases the
QCD-Compton process $\gamma q \rightarrow \gamma q (gq)$, and the photon-gluon
fusion (PGF) creation of a light $q \bar{q}$ pair. An
additional selection $\xbj <0.05$ allows to restrict the data to a region
where $\Delta q/q$ is closed to 0, and therefore the contribution to the
asymmetry from background processes (LO and QCD-Compton) can be neglected. With
this assumption we can relate the
%, except for events with $\xbj <0.05$, where the
%contribution to the measured asymmetry from the LO DIS and QCD-Compton is small. 
%The 
%
preliminary virtual photon-deuteron asymmetry $A^{\gamma^* d}$ resulting for
high-$p_T$ pair production from the 2002-2003 data at $Q^2>1$ (GeV$/c$)$^2$ ($A^{\gamma^* d}
= -0.015\pm0.080\,\mrf{(stat)} \pm0.013\,\mrf{(syst.)}$) to \DGG:
%Neglecting contributions to the asymmetry from other processes, we can relate
%the measured asymmetry to \DGG: 
$$
A^{\gamma^*d} = \frac{A_{LL}^{\mu d \rightarrow h h X}}{D} \approx \biggl\langle
  \frac{\hat{a}^{PGF}_{LL}}{D} \biggl\rangle \biggl\langle \frac{\Delta \mrf{G}}{\mrf{G}}
  \biggl\rangle \frac{\sigma^{PGF}}{\sigma^T} 
$$ 
where $\hat{a}_{LL}$ is the analyzing power of the process at the partonic
level, $D$ is the depolarization factor and $\sigma^{PGF}/\sigma^T$ is the
fraction of PGF processes in the selected sample.  For COMPASS we have estimated, from the
Monte-Carlo of the experiment, using LEPTO\cite{lepto} as generator, $\langle
\hat{a}_{LL}/D \rangle=-0.74 \pm 0.05$ and $\sigma^{PGF}/\sigma^T=0.34\pm 0.07$,
allowing the extraction of 
$\DGG = 0.06 \pm 0.31\, (\mrf{stat.}) \, \pm 0.06 \, (\mrf{sys.})$
at a mean gluon momentum fraction $x_g = 0.13$.

For the whole period 2002-2004 one can determine \DGG\ with an accuracy
$\simeq$~0.17 for events with $Q^2 >1$ (GeV$/c$)$^2$, while allowing for all $Q^2$
gives a statistical error of 0.05. The preliminary result from high-$p_T$ are
shown in figure~\ref{f:dg}(b), together with the measurements from
HERMES\cite{hermes-dg} and SMC\cite{smc-dg} .
\subsection{Collins and Sivers Effects}
The chirally-odd transversity distributions \dtq can be accessed, as suggested by
Collins\cite{collins}, in semi-inclusive interactions of
leptons off transversely polarized nucleons in combination with the chirally-odd
fragmentation function $\Delta D^h_q(z,p_T)$.  At first order the measured
azimuthal asymmetry $A\mrf{_{Col}}$ can be written as:
\[
\mrf{A_{Col}}=\frac{\sum_a e^2_a \cdot \Delta_T q_a (x) \cdot \Delta D^h_a(z, p_T)}{
  \sum_a e^2_a \cdot q_a(x) \cdot D^h_a(z)} = \frac{1}{\sin \Phi\mrf{_C} D\mit{_{NN}}
  f}\,\cdot\, \frac{N^+-N^-}{N^++N^-}
\]
A different mechanism may also give rise to an asymmetry in the scattering of
leptons off transversely polarized nucleon. Accounting for an
intrinsic momentum $k_T$ dependence of the quark distribution in the nucleon
$\Delta^T_0 q_a(x,k^2_T)$ may induce an azimuthal asymmetry (Sivers
effect\cite{sivers}) in: 
\[
\mrf{A_{Siv}}=\frac{\sum_a e^2_a \cdot \Delta^T_0 q_a(x,k^2_T)  \cdot D^h_a(z)}{
  \sum_a e^2_a \cdot q_a(x) \cdot D^h_a(z)} = \frac{1}{\sin \Phi\mrf{_S} D\mit{_{NN}}
  f}\,\cdot\, \frac{N^+-N^-}{N^++N^-}
\]
To measure transverse asymmetries, COMPASS has taken SIDIS data with the $^6$LiD
polarized orthogonally to the incoming $\mu$ momentum for about 20\% of the
running time.

\begin{figure}\begin{center}
\includegraphics[width=\columnwidth]{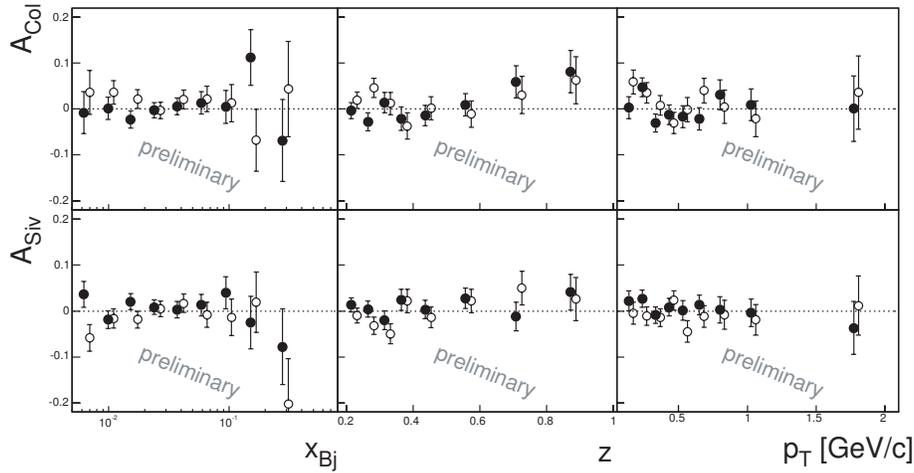} 
\end{center}\vspace{-2ex}
\caption{Collins (top) and Sivers (bottom) asymmetries as a function of \xbj,
  $z$ and $p_T$ for positively (open triangles) and negatively (closed squares)
  charged hadrons.}\label{f:collins}\end{figure}

Figure~\ref{f:collins} shows preliminary results of the first ever measured
single hadron Collins asymmetry on a deuteron target, separately for
positive and  negative hadrons produced in the struck quark fragmentation as a
function of \xbj, $z=E_h/(E_\mu-E_{\mu'})$ and the transverse momentum of the
hadron $p_T$ (first row). The Sivers asymmetry, also on a deuteron target is
shown in the second row.
The small values of the Collins asymmetries at all $x$ might imply either a
cancellation between the proton and the neutron asymmetries, or a small Collins
effect in the fragmentation $\Delta D^h_q$.  Also the Sivers asymmetry is small
and, with these statistical errors, compatible with zero, which may indicate a
small value of $\Delta^T_0 q$ in the covered $x$ range. These results are based
on 2002 run only; the full analysis of the 2002-2004 data will allow to reduce
the statistical errors by a factor of 2.

Other promising channels, such as the asymmetries in the two hadrons
frag\-men\-ta\-tion\cite{rainer}, inclusive vector mesons production, and $\Lambda$
productions are also under study and will give soon new important insight in the
nucleon spin structure.
\section{Conclusion}
The CERN COMPASS experiment is on the floor, taking data, since 2002. 
Many novel detectors and LHC techniques were integrated in the experiment and are  
performing according to expectations. The new and modern analysis system has
also required large efforts within the collaboration, but is now running at full gears
and very interesting results are coming.

First physics results have been produced and have been shown here and in the
parallel sessions of this conference, while many others will appear in the near
future. The first three years of data takin had shown that COMPASS has the
complete potentiality to fully perform its broad physics programme, allowing
CERN to bringing new pieces of information to the spin puzzle. After the
accelerator technical stop of 2005, COMPASS will run till 2010.

\end{document}

%% file: standard.tex
%-*-Mode: TeX; Mode: Auto-Fill; fill-column: 80; -*-
%
%--- definizioni con variabili:
%
\newcommand{\bfm}[1]{\mbox{\boldmath $ #1 $}}
\newcommand{\mrf}[1]{\mbox{$\mathrm{#1}$}}
\newcommand{\mem}[1]{\mbox{$\mathit{#1}$}}
\newcommand{\mat}[1]{\mbox{$#1$}}
%
%--- symboli di fisica:
%
\newcommand{\dds}{\mat{\mrf{\sigma(\Omega)}}}
\newcommand{\dsol}{\mat{\mrf{d} \sigma / \mrf{d} \Omega}}
\newcommand{\dstl}{\mat{\mrf{d} \sigma / \mrf{d} t}}
\newcommand{\dsof}{\mat{\frac{\mrf{d} \displaystyle{\sigma}}{\mrf{d}
\displaystyle{\Omega}}}}
\newcommand{\dsil}{\mat{(\mrf{d} \sigma / \mrf{d} \Omega)_i}}
\newcommand{\dsif}{\mat{\left(\frac{\mrf{d} \sigma}{\mrf{d} \Omega} \right)}}
\newcommand{\apw}{\mat{A_{0n}}}
\newcommand{\dnn}{\mat{D_{0n0n}}}
\newcommand{\knn}{\mat{K_{n00n}}}
\newcommand{\pio}{\mrf{\pi}}
\newcommand{\piz}{\mrf{\pi^{0}}}
\newcommand{\pip}{\mrf{\pi^{+}}}
\newcommand{\pim}{\mrf{\pi^{-}}}
\newcommand{\ppm}{\mrf{\pi^{\pm}}}
\newcommand{\ppi}{\mrf{\pi p}}
\newcommand{\elp}{\mrf{e^{+}}}
\newcommand{\elm}{\mrf{e^{-}}}
\newcommand{\epm}{\mrf{e^{\pm}}}
\newcommand{\g}{$\gamma$}
\newcommand{\N}{\mrf{N}}
\newcommand{\Z}{$Z$}
\newcommand{\A}{$A$}
\newcommand{\NB}{\mrf{\bar{N}}}
\newcommand{\NN}{\mrf{NN}}
\newcommand{\NNNN}{\mrf{NN \rightarrow NN}}
\newcommand{\NBNNBN}{\mrf{\bar{N}N \rightarrow \bar{N}N}}
\newcommand{\NBNpipi}{\mrf{\bar{N}N \rightarrow \pi\pi}}
\newcommand{\NBNpNBN}{\mrf{\bar{N}N \rightarrow \pi \bar{N}N}}
\newcommand{\NBN}{\mrf{\bar{N}N}}
\newcommand{\NBA}{\mrf{\bar{N}A}}
\newcommand{\pbar}{\mrf{\bar{p}}}
\newcommand{\pbarp}{\mrf{\bar{p}p}}
\newcommand{\pbarn}{\mrf{\bar{p}n}}
\newcommand{\Nbar}{\mrf{\bar{N}}}
\newcommand{\nbar}{\mrf{\bar{n}}}
\newcommand{\p}{\mrf{p}}
\newcommand{\n}{\mrf{n}}
\newcommand{\nbarA}{\mrf{\bar{n}A}}
\newcommand{\nbarn}{\mrf{\bar{n}n}}
\newcommand{\nbarp}{\mrf{\bar{n}p}}
\newcommand{\pbppbp}{\mrf{\bar{p}p \rightarrow \bar{p}p}}
\newcommand{\pbpnbn}{\mrf{\bar{p}p \rightarrow \bar{n}n}}
\newcommand{\pbpnbnp}{\mrf{\bar{p}p \rightarrow \pi^0 \bar{n}n}}
\newcommand{\nbnpbp}{\mrf{\bar{n}n \rightarrow \bar{p}p}}
\newcommand{\pppp}{\mrf{pp \rightarrow pp}}
\newcommand{\npnp}{\mrf{np \rightarrow np}}
\newcommand{\nppn}{\mrf{np \rightarrow pn}}
\newcommand{\nbch}{\mrf{\bar{n}A \rightarrow \bar{p}X}}
\newcommand{\pbpai}{\mrf{\bar{p}p \rightarrow \pi^+ \pi^-}}
\newcommand{\pbk}{\mrf{\bar{p}p \rightarrow K^+ K^-}}
\newcommand{\piN}{\mrf{\pi  N}}
\newcommand{\piNN}{\mrf{\pi NN}}
\newcommand{\pNb}{\mrf{\pi \bar{N}}}
\newcommand{\pNBN}{\mrf{\pi \bar{N}N}}
\newcommand{\ppmnn}{\mrf{\pi^{\pm}NN}}
\newcommand{\ponn}{\mrf{\pi^{0}NN}}
\newcommand{\piNpiN}{\mrf{\pi N \rightarrow \pi N}}
\newcommand{\pippin}{\mrf{\pi^+ p \rightarrow \pi^0 n}}
\newcommand{\puppup}{\mrf{\pi^+ p \rightarrow \pi^+ p}}
\newcommand{\pdppdp}{\mrf{\pi^- p \rightarrow \pi^- p}}
\newcommand{\dsdo}{${d\sigma \over d\Omega}$}
\newcommand{\tu}{$\tau_1$}
\newcommand{\td}{$\tau_2$}
\newcommand{\tr}{$\tau_3$}
\newcommand{\tap}{$\tau_+$}
\newcommand{\tam}{$\tau_-$}
\newcommand{\iu}{$I_1$}
\newcommand{\id}{$I_2$}
\newcommand{\ir}{$I_3$}
\newcommand{\ih}{$I_+$}
\newcommand{\il}{$I_-$}
\newcommand{\su}{$\sigma_1$}
\newcommand{\sd}{$\sigma_2$}
\newcommand{\st}{$\sigma_3$}
\newcommand{\sih}{$\sigma_+$}
\newcommand{\sil}{$\sigma_-$}
\newcommand{\ST}{$S_{12}$}
\newcommand{\QL}{$Q_{12}$}
\newcommand{\biq}{$\bfm{I}^2$}
\newcommand{\thf}{$\theta_{cm}^f$}
\newcommand{\tsc}{$\theta_{sc}$}
\newcommand{\tcm}{$\theta_{cm}$}
\newcommand{\tne}{$\theta_{n}$}
\newcommand{\tnb}{$\theta_{\bar{n}}$}
\newcommand{\paz}{$\phi_{\bar{n}}$}
\newcommand{\zer}{$0^{\circ}$}
\newcommand{\gr}{\mat{^{\circ}}}
\newcommand{\SQ}{$Q^2$}
\newcommand{\fpnn}{$f_{nn \pi^0}$}
\newcommand{\fppp}{$f_{pp \pi^0}$}
\newcommand{\fpc}{$f_{pn \pi^+} f_{np \pi^-}$}
\newcommand{\fqpnn}{$f_{nn \pi^0}^2$}
\newcommand{\fqppp}{$f_{pp \pi^0}^2$}
\newcommand{\fq}{$f^2$}
\newcommand{\gq}{$g^2$}
\newcommand{\fqc}{$f^2_c$}
\newcommand{\fqo}{$f^2_0$}
\newcommand{\gqc}{$g^2_c$}
\newcommand{\gqo}{$g^2_0$}
\newcommand{\thcm}{$\theta_{cm} = 90^{\circ}$}
%
%--- rivelatori
\newcommand{\ANU}{ANC$_1$}
\newcommand{\AND}{ANC$_2$}
\newcommand{\ANT}{ANC$_3$}
\newcommand{\bo}{B$_0$}
\newcommand{\bu}{B$_1$}
\newcommand{\bou}{B$_0 \cdot$B$_1$}
\newcommand{\xbj}{\mat{x}\mrf{_{Bj}}}
\newcommand{\y}{$y$}
\newcommand{\z}{$z$}
\newcommand{\xz}{$xz$}
\newcommand{\yz}{$yz$}
\newcommand{\thx}{$\theta_x$}
\newcommand{\thy}{$\theta_y$}
\newcommand{\cfft}{\mrf{C_4F_{10}}}
%
%--- unita' di misura
\newcommand{\mn}{\mrf{\mu m}}
\newcommand{\mm}{\mrf{mm}}
\newcommand{\mq}{\mrf{mm^2}}
\newcommand{\mc}{\mrf{mm^3}}
\newcommand{\cq}{\mrf{cm^2}}
\newcommand{\cmc}{\mrf{cm^3}}
\newcommand{\um}{\mrf{\mu m}}
\newcommand{\MO}{M$\Omega$}
\newcommand{\mom}{\mrf{MeV}\mat{/c}}
\newcommand{\mee}{\mrf{MeV}\mat{_{ee}}}
%
%--- accenti italiani:
% minuscole
\newcommand{\est}{\`e }
\newcommand{\nea}{n\`e }
\newcommand{\piu}{pi\`u }
\newcommand{\puo}{pu\`o }
\newcommand{\per}{perch\'e }
\newcommand{\percio}{perci\`o }
\newcommand{\nonche}{nonch\`e }
\newcommand{\poi}{poich\'e }
\newcommand{\cio}{ci\`o }
\newcommand{\cioe}{cio\`e }
\newcommand{\cosi}{cos\`{\i} }
\newcommand{\pero}{per\`o }
\newcommand{\gia}{gi\`a }
% maiuscole
\newcommand{\Est}{\`E }
\newcommand{\Piu}{Pi\`u }
\newcommand{\Puo}{Pu\`o }
\newcommand{\Per}{Perch\'e }
\newcommand{\Poi}{Poich\'e }
\newcommand{\Cio}{Ci\`o }
\newcommand{\Cioe}{Cio\`e }
\newcommand{\Cosi}{Cos\`{\i} }
\newcommand{\Pero}{Per\`o }
\newcommand{\Gia}{Gi\`a }
%
% --- spaziature
%
\newcommand{\hex}{\hspace{0.5ex}}
\newcommand{\hem}{\hspace{0.5em}}
\newcommand{\hcm}{\hspace{0.5cm}}
\newcommand{\ucm}{\hspace{1.0cm}}